\documentclass[traditabstract]{aa}
\usepackage[varg]{txfonts}
\usepackage{graphicx}
\usepackage{graphics}
\usepackage[T1]{fontenc}
\usepackage[utf8]{inputenc}
\newcommand{\td}[1]{\, \mbox{d} #1 \,}
\newcommand{\intl}{\int\limits}

\newcommand{\blob}{{blob}}

\newcommand{\pcjet}{{pc-jet}}
\newcommand{\extendedjet}{{extended jet}}
\begin{document}
\title{The extended jet of AP Librae: Origin of the very high-energy $\gamma$-ray emission?}
\author{Michael Zacharias\thanks{E-mail: m.zacharias@lsw.uni-heidelberg.de} \and Stefan J. Wagner}
\institute{Landessternwarte, Universit\"at Heidelberg, K\"onigstuhl, D-69117 Heidelberg, Germany}
\date{Received ? / accepted ? }
\abstract
{The low-frequency peaked BL Lac object (LBL) AP Librae exhibits very-high-energy (VHE, $E>100$GeV) $\gamma$-ray emission and hosts an extended jet, which has been detected in radio and X-rays. The jet X-ray spectral index implies an inverse Compton origin. These observations are unusual for LBLs calling for a consistent explanation of this extraordinary source. The observationally constrained parameters necessary to describe the core emission within the standard one-zone model are unable to explain the broad-band spectrum, even if observationally unconstrained external photon fields are taken into account. We demonstrate that the addition of the extended jet emission successfully reproduces the total spectral energy distribution. In particular, the VHE radiation is produced in the $>100\,$kpc long extended jet via inverse Compton scattering of cosmic microwave background photons by highly relativistic electrons. We present several ways to test this theory. The extended jet is weakly magnetized ($B_0 = 2.5\,\mu$G), while its minimum and maximum electron Lorentz factors are $\gamma_{min}=60$ and $\gamma_{max}=5\times 10^6$, respectively. The electron spectral index is $s=2.6$. These parameters are comparable to parameters of other blazars with extended X-ray jets dominated by inverse Compton scattering.}
%
\titlerunning{The extended jet of AP Librae}
\authorrunning{M. Zacharias \& S.J. Wagner}
\maketitle
%
%
\section{Introduction}
Blazars are a class of active galactic nuclei (AGN), where the jet is closely aligned with the line of sight (\cite{br74}) according to the AGN unification scheme. This amplifies the non-thermal jet emission due to relativistic beaming.

Blazars exhibit a two-component spectral energy distribution (SED), where the low-energy component is attributed to synchrotron emission of relativistic electrons, and the high-energy component originates from inverse Compton (IC) scattering of ambient soft photons by the same electron population. Such soft photon fields include the synchrotron photons (synchrotron self-Compton, SSC), photons from the accretion disk (IC/disk), and thermal photon fields, such as the broad-line region, or the dusty torus (IC/DT). Soft photon fields are also provided by the galactic starlight, or from the cosmic microwave background (IC/CMB). A review of these leptonic models is given by, e.g., \cite{b07}. In hadronic models, proton interactions and resulting pion decay and leptonic cascade processes are responsible for the high-energy emission (e.g.,\cite{bea13,ws15,cea15}). These processes usually require greater jet powers and much higher magnetic field strengths than the leptonic models and are therefore under criticism (e.g.,\cite{zb15,pd15}). In this work we will concentrate on leptonic processes.

BL Lac objects can be characterized by the peak position of their synchrotron component in the SED. Low-energy peaked BL Lac objects (LBLs) exhibit the synchrotron maximum below $10^{14}\,$Hz, while intermediate-energy peaked BL Lac objects (IBLs) peak between $10^{14}\,$Hz and $10^{15}\,$Hz, and high-energy peaked BL Lac objects (HBLs) peak above $10^{15}$Hz. 

The blazar AP Librae, located at redshift $z=0.0486$ and at the coordinates $\mbox{R.A.} = 15^h 17^m 41.8^s$, $\mbox{Dec.} = -24^{\circ} 22^{\prime} 19.5^{\prime\prime}$ (J2000), exhibits a strictly increasing X-ray energy spectrum. In combination with the lack of optical emission lines it is classified as an LBL. However, some observational features of this source do not fit into this category.

Since LBLs exhibit their synchrotron peak frequency below $10^{14}\,$Hz and the X-rays originate from the IC process, the maximum electron Lorentz factor in the electron energy distribution does not significantly exceed $\sim 10^4$ for reasonable values of the magnetic field strength on the order of $\sim 0.1\,$G. Hence, one would not expect very high-energy $\gamma$-ray (VHE, $E>100\,$GeV) emission from these objects. Nevertheless, AP Librae has been clearly detected by observations with the H.E.S.S. telescope array (\cite{hea15}) and the SED extends to energies of a few TeV. Currently, AP Librae is the only LBL listed in the TeVCat\footnote{http://tevcat.uchicago.edu/}, a catalog that gathers all sources detected above $100\,$GeV. Despite selection biases, this makes AP Librae an exceptional source.

High-resolution X-ray observations led to the detection of extended X-ray jets in many AGN. However, owing to the small viewing angle, it is surprising to observe extended X-ray emission in blazars. An extended X-ray jet has been detected in AP Librae by \cite{kwt13}, which has thus become one of only six BL Lac objects listed in the X-JET database.\footnote{http://hea-www.cfa.harvard.edu/XJET/} In three of these objects the X-ray emission is attributed to synchrotron emission, while in AP Librae the X-ray photon index of $1.8\pm 0.1$ suggests IC emission. The X-ray morphology of AP Librae's jet exactly follows the radio morphology as observed with the VLA. This suggests a common origin of the radio and X-ray radiation. The detection of the extended X-ray jet further demonstrates AP Librae's peculiar state.

The detection in VHE implies an extremely broad high-energy component spanning 10 orders of magnitude in frequency. Owing to the observed cut-off of the synchrotron emission below the X-rays and other observational constraints, the broad high-energy component cannot be explained in the usual one-zone blazar model. Even by considering possible, observationally unconstrained extensions of the one-zone model (such as external photon fields like the BLR or the dusty torus), the VHE emission cannot be explained.

In this work, we explore the possibility that the extended jet contributes to the total SED in the VHE $\gamma$-ray regime. In fact, our model explains the VHE emission as originating mostly from the extended jet by IC/CMB radiation. This model has also been suggested, for example, for the extreme BL Lac object 1ES 1101-232 (\cite{bdf08}). However, the observational basis in 1ES 1101-232 is less well constrained than in AP Librae.

The paper is organized as follows. First, we present the data set in section \ref{sec:data}, and give a special emphasis to resolved structures on arcsec and milliarcsec scales. Section \ref{sec:fit} contains the actual modeling, where we confirm the failure of the one-zone model, and then explain the SED by modeling the resolved structures. The details of the numerical code are given in appendix \ref{app:code}. In section \ref{sec:dis} we discuss the power requirements of our model, potential observations to verify our theory, and compare AP Librae to similar sources. Our conclusions are presented in section \ref{sec:con}.

We use a standard cosmology with $H_0 = 70\,$km s$^{-1}$Mpc$^{-1}$, implying a luminosity distance of AP Lib of $d_L = 6.48\times 10^{26}\,$cm. The corresponding angular scale equals $1\,$arcsec$=0.95\,$kpc. 

Since the observations in different energy bands exhibit different resolutions, we invoke the following nomenclature. We refer to {\it core} for any data that is unresolved irrespective of the scale. Resolved structures are referred to as the {\it jet}. The compact region, which is typically used to explain a blazar SED within the ``one-zone'' model, is referred to as the {\it blob}.

We will model the jet from its base out to arcsec scales. Owing to observational constraints, it is divided into two parts. With increasing distance from the center these are the {\it pc-jet} on milliarcsec scales, and the {\it extended jet} on arcsec scales. This choice will be discussed in detail in sections \ref{sec:mojave} and \ref{sec:multi}.

%
%
\section{Structure and SED of AP Librae} \label{sec:data}
\begin{figure}
\centering
\includegraphics[width=0.48\textwidth]{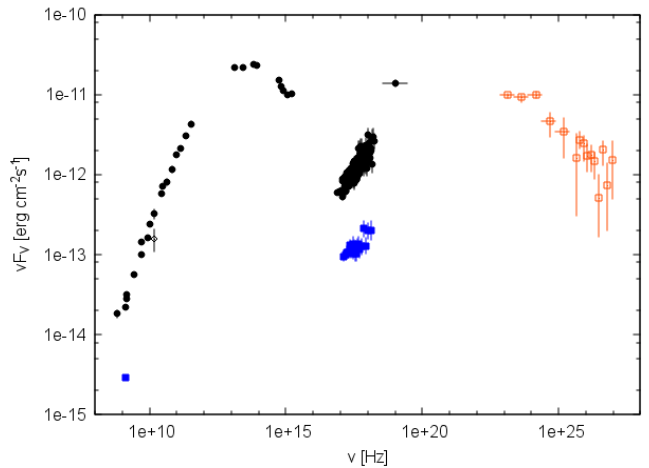}
\caption{SED of AP Librae. Data attributed to the core on arcsec scales is shown as black dots. Blue squares correspond to the extended jet data. The open diamond at $15$GHz is the flux from component 0 of the MOJAVE observations. The orange open squares are the $\gamma$-ray data from FERMI-LAT and the H.E.S.S. experiment where no discrimination between core and jet is possible.}
\label{fig:spec}
\end{figure}

AP Librae has been extensively observed in all energy bands, but most of the energy bands have not been observed simultaneously. This is unfortunate, since AP Lib is known to be highly variable at optical frequencies (e.g., \cite{wea88,cea91}). We assume that the chosen data represents an average state of the source.

The SED is shown in Fig. \ref{fig:spec}. The black dots represent data attributed to the core region on arcsec scales, while the blue squares correspond to the extended jet data. The orange open squares mark the $\gamma$-ray data, where no discrimination concerning the angular extent is possible. The details of each energy range are presented below.
%
\subsection{Radio}
The radio data points presented in Fig. \ref{fig:spec} of the core emission are derived from VLA observations (\cite{cea99}), from \cite{kea81}, from the MOJAVE catalog (\cite{lea13}), and from the PLANCK Legacy Archive (PR1).\footnote{http://www.cosmos.esa.int/web/planck/pla} At $\sim 20\,$GHz the SED exhibits a break from a $\nu^{1}$ to a softer $\nu^{0.8}$ spectrum. 

The data point of the extended jet at $1.36\,$GHz is taken from \cite{cea99}. The extended jet emerges to the southeast. It bends after $\sim 10\,$arcsec to the northeast and has been traced out to $40\,$arcsec.

\subsubsection{MOJAVE observations} \label{sec:mojave}

On milliarcsec scales ($1\,$milliarcsec corresponds roughly to $1\,$pc) the jet is resolved by the MOJAVE VLBI observations at $15\,$GHz. We show the average flux (observations between 1998 and 2013) of the milliarcsec core ({\it component 0} as given in the MOJAVE catalog) as the open diamond in Fig. \ref{fig:spec}. Component 0 is variable, but fluxes have not varied by more than a factor 2 within the 15 years of observations. The vertical error bar displays the temporal variation within the data set. The black filled data point at $15\,$GHz is the average of the total flux of each MOJAVE observation, including both the component 0 and the resolved milliarcsec jet. This suggests that all black radio points in the SED are likely to contain flux from the actual blazar zone and an extended region of $\gtrsim$pc scales. 

\cite{lea13} discovered three knots which they were able to follow over a few years. These resulted in the measurement of a maximum apparent speed of the jet of $\beta_{app} = v_{app}/c = 6.39$.

The MOJAVE data indicates a continuous jet on $10\,$milliarcsec scales emerging to the south in all observations since 2007.\footnote{http://www.physics.purdue.edu/astro/MOJAVE/sourcepages/1514-241.shtml} Beyond these scales the jet bends to the southeast, which corresponds to the direction of the extended jet on arcsec scales.

%
\subsection{Infrared, optical, UV}
In the infrared, optical, and UV bands we use the data presented in \cite{kwt13} and \cite{hbs15}.

The infrared data is taken from the WISE mission, while the optical and UV data points are derived from Swift-UVOT observations. All data was taken in 2010 and is corrected for Galactic absorption, and for the contribution of the host galaxy.

As stated above, AP Librae is highly variable at optical frequencies. This makes it difficult to determine the average flux state. \cite{kwt13} present R- and B-band data taken with the ATOM telescope located at the H.E.S.S. site in Namibia. Their data is an average over the time range of the 2FGL catalog (data taken between 2008 and 2010; \cite{nea12}). The data points are compatible with the Swift-UVOT measurement.

Despite observations with large optical telescopes and the Hubble Space Telescope, no hint of the extended jet has been observed at optical wavelengths so far (e.g.,\cite{sea00}). This is due to the fact that the host galaxy is very bright in the optical even several arcsec from the core (values for the half-light radius in the literature range from $3.7\,$arcsec (\cite{sea00}) up to $6.72\,$arcsec (\cite{pea02})). Hence, the jet emission could be hidden by the starlight.
%
\subsection{X-rays}
AP Librae was observed with the Chandra satellite in 2003. The high spatial resolution of Chandra allowed the detection of the kpc-jet in the X-ray regime by \cite{kwt13}, who also derived the spectral indices for both the core and the jet. The filled black points in Fig. \ref{fig:spec} belong to the core region of the blazar, while the blue filled squares show the data of the extended jet. The photon index of the core emission equals $1.58\pm 0.04$, which is consistent with AP Librae being an LBL and the X-ray radiation originating from IC emission. The extended jet photon index of $1.8\pm 0.1$ indicates that the X-ray jet is also IC dominated.

The X-ray jet detected by Chandra is spatially coincident with the extended jet at $1.36$GHz and emerges to the southeast (\cite{kwt13}). Beyond the bend in the radio, the X-ray emission is much weaker than before the bend and fades rapidly.

The hard X-ray data point above the Chandra core spectrum in Fig. \ref{fig:spec} is taken from the Palermo 100-month Swift-BAT catalog.\footnote{http://bat.ifc.inaf.it/} The photon index of the Swift-BAT measurement is given as $1.69\pm 0.14$. This is consistent with the Chandra core spectrum. The flux level has been steady within errors over all the Palermo Swift-BAT catalogs (39 months, 54 months, 66 months, and 100 months). Hence, this value is not particularly influenced by potential flaring events in this energy band. 

The consistency of the Chandra core spectrum and the Swift-BAT spectrum give confidence that the Chandra observations were conducted during a quiescent period. If this were not the case, then the quiescence spectrum would be harder than the Chandra core spectrum, which would complicate the modeling beyond the current constraints (see section \ref{sec:fit}). A Swift-XRT spectrum from $2011$ exhibits a slightly higher flux than the Chandra core spectrum, which is consistent with the assumption of the Chandra core spectrum representing an average spectral state (\cite{kwt13}).
%
\subsection{Gamma rays} 
The data at $\gamma$-ray energies is taken from \cite{hea15}, and include measurements from the FERMI satellite in the high-energy (HE, $E>100\,$MeV) band and from the H.E.S.S. experiment in the VHE band. A distinction between the core and the extended jet is not possible at these energies owing to the lack of the necessary spatial resolution.

The Fermi data points are taken from the long quiescence period in HE, where no variability is detected. \cite{hea15} also report an outburst of AP Librae in the Fermi-LAT energy range with a maximum flux increase of a factor 3.5 above the quiescence level. The entire flaring episode with a relatively long decay lasted for about 100 days before the flux returned to pre-flare levels. We examined this flare to ascertain that it is not caused by confusion with a luminous GeV flare in the nearby quasar PKS 1510-089 that is coincident with the AP Librae outburst.

The H.E.S.S. data was taken in 2010 and 2011. No indication of flux variability at the VHE energy range has been published. No H.E.S.S. result is reported for the Fermi flare period in 2013. 

We de-absorbed the spectral points using the EBL model of \cite{fea08}, giving an intrinsic flux roughly $32\%$ higher than the absorbed flux at $1$ TeV.

%
%
\section{Modeling AP Librae} \label{sec:fit}
The observed characteristics of the extraordinary source are studied using the leptonic steady-state code of \cite{bea13}. Details are presented in Appendix \ref{app:code}. There we also describe our addition to the code used to model the extended component. This is assumed to be homogeneous. It is treated as an ensemble of identical subunits, and the combined emission of the subunits gives the total flux of the extended component.

Below we first consider constraints on the free parameters, which can be deduced from the observations. These constraints and our modeling rule out the one-zone model. We explore whether the SED can be reproduced by considering the emission of the entire jet. It will be divided into two parts. These are the \pcjet, which describes the structures on milliarcsec scales resolved by the MOJAVE observations, and the \extendedjet{} on kpc scales, which is probed by radio and X-ray measurements.

%
\subsection{Constraints} \label{sec:cons}
The most obvious difficulty for modeling AP Librae is the broad inverse Compton component in combination with the Swift-BAT measurement at hard X-rays. The latter is a long-term average and not significantly influenced by potential outbursts. Since it coincides with an extrapolation of the Chandra core spectrum, the IC component (which is responsible for the X-ray emission) must exhibit a hard power law over two orders of magnitude in energy. The absence of any curvature in this energy range restricts the minimum electron Lorentz factor $\gamma_{min}$ to values above $10^2$ assuming a standard electron spectral index $s=2$, as is discussed in detail in Appendix \ref{app:ssc}.

The hard X-ray energy flux exceeds the flat Fermi-LAT level. This indicates that the peak frequency of the inverse Compton component is located below the Fermi energy range. In turn, this requires that the maximum electron Lorentz factor $\gamma_{max}$ not be much larger than $10^4$ in order to meet the fluxes at $\gamma$-rays.

These constraints on the electron distribution determine the synchrotron component in most characteristics. The magnetic field $B_0$ is limited to $0.01 - 0.1\,$G by the optical and UV measurements.

Generally, using a self-consistent model implies that two out of five parameters of the broken power law of the electron distribution function are determined by other parameters. The break Lorentz factor $\gamma_{br}$, as given in Eq. (\ref{eq:gammab}), depends sensitively on the source radius and the magnetic field. Furthermore, the slope of the second power law is restricted to values $s+1$, as given in Eqs. (\ref{eq:fastcooling}) and (\ref{eq:slowcooling}). Since the standard Fermi acceleration mechanisms provide spectral indices on the order of $s\sim 2$, the parameter space is already much reduced.

From the measurement of the maximum apparent speed of the radio knots on pc-scales, the values of the Doppler factor $\delta_b$ and the angle between the inner jet and the line of sight $\vartheta_{obs}$ can be deduced following Eqs. (2.71) and (2.72) of \cite{b12}. The bulk Lorentz factor $\Gamma_b$ is not empirically constrained. Throughout the paper we assume a non-decelerating jet and set $\Gamma_b = 10$ in all components. This value is chosen to avoid the so-called Doppler factor crisis in blazars (e.g.,\cite{hs06}). This gives $\delta_b = 17.8$ and $\vartheta_{obs}=2^{\circ}$ in the inner regions of the jet. Our conclusions are insensitive to this assumed value.

Using these constraints, we give the modeling results below. The input parameters for the models are given in Table \ref{tab:inputcom}. The results of the combined model are listed in Table \ref{tab:resultcom}. 

%
\subsection{One-zone model} \label{sec:ozm}
\begin{table}
\caption{Input parameters for the models. The definitions of the parameters are given in the text and in Appendix \ref{app:code}.}
\begin{tabular}{lccc}
				  & Blob 		& pc-jet 		& Extended jet \\
	\hline
	$L_{inj}$ [erg/s] 	  & $2.4\times 10^{44}$ & $5.0\times 10^{40}$ 	& $6.0\times 10^{42}$ \\
	$\gamma_{min}$ 		  & $1.6\times 10^2$ 	& $2.0\times 10^2$ 	& $4.0\times 10^1$ \\
	$\gamma_{max}$ 		  & $1.0\times 10^4$ 	& $1.0\times 10^4$ 	& $5.0\times 10^6$ \\
	$s$ 			  & $2.0$ 		& $2.0$ 		& $2.6$ \\
	$\eta_{esc}$ 		  & $45$		& $40$	 		& $15$ \\
	$B_{0}$ [G] 		  & $0.04$ 		& $5.0\times 10^{-3}$ 	& $2.3\times 10^{-6}$ \\
	$R$ [cm] 		  & $9.0\times 10^{15}$ & $3.0\times 10^{18}$ 	& $1.5\times 10^{21}$ \\
	$z_{0}$ [pc] 		  & $0.3$ 		& $1.0$ 		& $100$ \\
	$\Gamma_{b}$ 		  & $10$ 		& $10$ 			& $10$ \\
	$\vartheta_{obs}$ [deg]	  & $2.0$ 		& $2.0$ 		& $4.0$ \\
	$L_d$ [erg/s] 		  & $1.3\times 10^{44}$	& - 			& - \\
	$\tau_{DT}$ 		  & $0.01$		& - 			& - \\
	$r_{DT}$ [pc] 		  & $1.8$		& - 			& - \\
	$T_{DT}$ [K] 		  & $5.0\times 10^{2}$	& - 			& - \\
	$l_{jet}^{\prime}$ [pc]	  & - 			& $10.0$		& $1.0\times 10^{4}$ \\
\end{tabular}
\label{tab:inputcom}
\end{table}
\begin{figure}
\centering
{\includegraphics[width=0.48\textwidth]{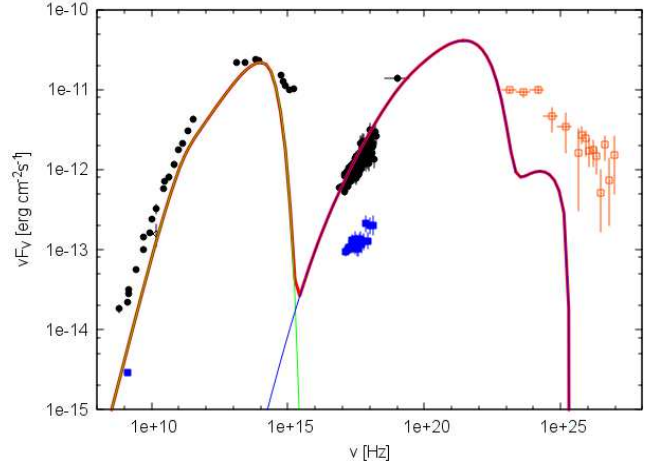}}
\caption{One-zone SSC model. Data points are as in Fig. \ref{fig:spec}. The lines correspond to synchrotron (green), SSC (blue), and the sum (thick red). }
\label{fig:ozssc}
\end{figure} 
\begin{figure}
\centering
\includegraphics[width=0.48\textwidth]{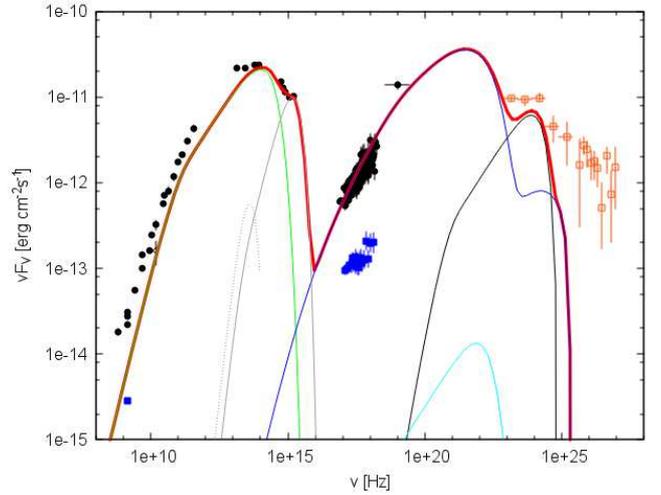}
\caption{One-zone model including thermal radiation fields and associated IC emission. Data points are as in Fig. \ref{fig:spec}. The lines correspond to synchrotron (green), accretion disk (gray), DT field (dotted gray), SSC (blue), IC/disk (cyan), IC/DT (black), and the sum (thick red). }
\label{fig:ozmodel}
\end{figure}

Here we present a one-zone model of AP Librae fulfilling the previously mentioned observational constraints. This model is described by the parameters given in the first column of Table \ref{tab:inputcom} and is shown in Fig. \ref{fig:ozssc}. We know from the MOJAVE observations that the radio data of the arcsec core contain emission from an extended region on milliarcsec scales. Thus, the one-zone model is not expected to fit the flux measurements in the radio (shown as black points in Fig. \ref{fig:spec}), but rather the unresolved flux on milliarcsec scales (shown as the open diamond). 

The model fits the X-ray part of the SED. The hard X-ray spectrum in reproduced by an electron distribution with $\gamma_{min} = 160$ for a standard spectral index of $s=2$. We refer to appendix \ref{app:ssc} for a detailed justification of this value.

The peak energy of the IC component is located in the range from $100\,$keV to $100\,$MeV. In order to fulfill the Fermi flux measurement at $100\,$MeV the SSC component must cut off below $100\,$MeV resulting in a maximum electron Lorentz factor of $10^4$. Thus, the electron distribution is quite narrow spanning less than two orders of magnitude in energy.

With these constraints on the electron distribution, the synchrotron component is determined except for the magnetic field. Choosing $B_0 = 0.04\,$G allows the fit of the flux of radio component 0, and the optical/NIR regime. The UV data is not fitted well by synchrotron emission alone (see below). 

While the resulting SSC emission fits the X-ray spectrum, it is unable to explain the $\gamma$-ray spectrum, owing to the excess at MeV energies compared to the Fermi flux level. Even the contribution of the second-order SSC emission, which is indicated in Fig. \ref{fig:ozssc} by the second hump of the blue curve at $\sim 10^{24}\,$Hz, cannot account for the $\gamma$-ray flux. 

Since AGN are associated with accreting systems, we expect AP Librae's supermassive black hole of mass $M_{bh} = 2.5\times 10^8\,$M$_{\odot}$ (\cite{wea05}) to be surrounded by an accretion disk. Since there is no evidence of the disk in the data, we can use the flux measurements in the optical/UV to set a maximum limit on the disk's luminosity. In Fig. \ref{fig:ozmodel} we add the thermal emission of a Shakura-Sunyaev-type disk (\cite{ss73}) with luminosity $L_D = 1.3\times 10^{44}\,$erg/s. The combination of the disk and the synchrotron component fits the Swift-UVOT flux measurements.

The disk emission is IC scattered by the electrons in the \blob. However, the outward direction of the relativistic motion of the \blob{}, and its distance from the accretion disk result in a minor contribution to the overall spectrum.

The accretion disk emission can be scattered by material surrounding the black hole on $\sim$pc scales. Since there is no indication of such a component in the data, the modeled flux level must be weak. The parameters of this potential photon field listed in Table \ref{tab:inputcom} suit a very weak dusty torus (DT). Following Eq. (\ref{eq:ublr}) the energy density of the DT field becomes $u_{DT} = 1.1\times 10^{-7}\,$erg/cm$^3$ implying a luminosity of $L_{DT} = 4.0\times 10^{42}\,$erg/s. This is more than $1$ order of magnitude below the luminosity of the accretion disk. This accounts for the non-detection of the DT field, as can be seen in Fig. \ref{fig:ozmodel}.

However, the DT field can be IC scattered by the \blob{} electrons. Owing to the geometry of the  photon field, it is blue-shifted in the frame of the \blob{} increasing both the maximum energy of the IC photon and the IC/DT flux. The parameters of the DT field were chosen such that the IC/DT component helps to explain the $\gamma$-ray emission. The flat Fermi level is represented well by the superposition of the SSC and the IC/DT components. The slight underestimate in Fig. \ref{fig:ozmodel} of the Fermi flux level could be accounted for by choosing a slightly higher energy density in the DT field. However, our choice is in anticipation of the contribution of the extended jet at these energies (see section \ref{sec:multi} and Fig. \ref{fig:extmodel}). 

The IC kinematics in the Klein-Nishina regime imply a firm cut-off at the observed frequency $\nu_{max} = \delta_b \gamma_{max} m_ec^2/h = 2.2\times 10^{25}\,$Hz. The IC component cuts off at this frequency, even though in most parts of the IC spectrum scatterings occur in the Thomson regime. Thus, the one-zone model cannot account for the VHE emission. Neither the disk nor the DT field can circumvent this conclusion. Invoking these fields to explain some parts of the SED is of minor importance compared to the failure of the one-zone model as a whole.

Synchrotron self-absorption can influence the synchrotron spectrum at low radio frequencies. Following Eq. (7.146) of \cite{dm09}, the turn-over frequency can be determined by
\begin{eqnarray}
 \nu_{ssa} = 2.65\times 10^{7} \delta_b \left( R_{15} n_e b^2 \right)^{1/3} \,\mbox{Hz}, \label{eq:nussa}
\end{eqnarray}
with the scalings $R = 10^{15}R_{15}\,$cm, $B_0 = b\,$G, and the electron density $n_e$. With the model parameters (see also Table \ref{tab:resultcom}), we derive $\nu_{ssa} = 7.6\times 10^{8}\,$Hz. Therefore, synchrotron-self absorption within the \blob{} does not affect the model.

The electrons cool mostly via the SSC channel with a cooling time of $\sim 140\,$days for the highest energy electrons. Such a long cooling time is expected, since the model aims to explain the time-averaged SED. Fast flaring episodes would require a time-dependent approach (e.g. \cite{zs12,cea14}).

We note that it would also be possible to reproduce the emission of the one-zone model with an electron distribution of lower $\gamma_{min}$ if a very hard spectral index of the electron population is assumed. Appendix \ref{app:ssch} illustrates that for an index of $s=1.5$ the minimum electron Lorentz factor $\gamma_{min}$ might be as low as $30$. Recent works on the theory of relativistic shocks (e.g.,\cite{cub12,sb12,s15,wea16}) claim that hard electron distributions are possible in certain circumstances. It is beyond the scope of this paper to investigate whether these circumstances are realized in AP Librae's jet, and it also cannot circumvent the necessity of additional components (c.f. Fig. \ref{fig:ozhard}). This clearly confirms the results of other authors (\cite{tea10,hbs15}) that the standard one-zone model fails to reproduce the broad-band SED of AP Librae, even if additional external photon fields with poor observational constraints are taken into account. This deficit of the one-zone model is independent of the precise details of assumptions beyond the observational constraints. 

Below we concentrate on contributions from the large-scale jet in order to successfully explain the SED.

%
\subsection{Jet model} \label{sec:multi}

As discussed in section \ref{sec:mojave}, the MOJAVE data resolves the jet emission on pc-scales. This demonstrates that a significant fraction of the total flux measured in radio at moderate angular resolution is confined in a region significantly smaller than the \extendedjet{} on arcsec scales, and yet much larger than the single zone described in section \ref{sec:ozm}. We refer to this region as the \pcjet. We take this region into account in addition to the one-zone model in order to explain the radio data. The one-zone model (or \blob) now represents a small region moving within the \pcjet{} region. 

\begin{figure}
\centering
\includegraphics[width=0.48\textwidth]{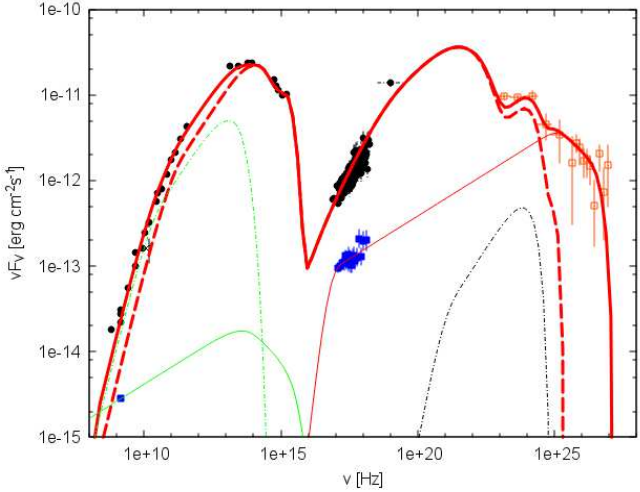}
\caption{Model including the extended components. Data points are as in Fig. \ref{fig:spec}. The lines correspond to the total emission of the \blob{} in Fig. \ref{fig:ozmodel} (thick, dashed red), synchrotron of the \pcjet{} (dash-dotted green), IC/DT of the \pcjet{} (dash-dotted black), synchrotron of the \extendedjet{} (solid green), IC/CMB of the  \extendedjet{} (solid red), and the sum of all components (thick, solid red). }
\label{fig:extmodel}
\end{figure}

The \pcjet{} is constrained by the condition that the combination of the two components should not exceed the radio and infrared data. The parameters of the \pcjet{} are summarized in the second column of Table \ref{tab:inputcom}, and the resulting fit is shown in Fig. \ref{fig:extmodel}. The combination of the \blob{} and the \pcjet{} matches the synchrotron component very well. Empirical parameters of the \pcjet{} are the projected length of $10\,$pc and a radius of $1\,$pc. The de-projected length of this part of the jet is $\sim 290\,$pc. The combination of the \blob{} and the \pcjet{} at $15\,$GHz equals the total flux from the MOJAVE observations.

Since SSC emission depends strongly on the local synchrotron photon density at any point within the jet, the SSC flux of the \pcjet{} is very low. The part of the jet immersed in the photon field of the accretion disk is very small since the photon density drops rapidly with height above the disk. These photon fields can be safely ignored. However, about $0.6\%$ of the \pcjet{} are immersed in the DT photon field. Even though the electron density is low in the jet, the large jet volume within $r_{DT}$ contains a number of electrons that is only a factor of $\sim 20$ smaller than the total number of electrons in the \blob. Hence, the IC/DT component of the \pcjet{} cannot be neglected, even though it does not contribute significantly to the overall flux (see Fig. \ref{fig:extmodel}).

In the \pcjet{} the electrons are cooled predominantly by synchrotron emission, and the highest energy electrons cool on a time scale of about $100\,$years. This cooling time is not long enough for electrons to cross the length of this jet region, and re-acceleration must take place on scales smaller than $30\,$pc.

In principle, the synchrotron emission of the \pcjet{} could be scattered in the much denser \blob{} medium. This effect can be neglected, however, because in the immediate vicinity of the \blob{} the synchrotron emission is very weak, and owing to the same bulk speeds there is no Doppler enhancement.

We assume the same observing angle for the \blob{} and the \pcjet{}, since there is no indication of a change in direction within the $10\,$milliarcsec jet. Beyond this distance the jet turns towards the southeast, which is the same direction as the jet on arcsec scales. The bend is assumed to correspond to an intrinsic change of the viewing angle, amplified by the foreshortening. The change is assumed to result in an observing angle of the \extendedjet{} of $4^{\circ}$. This reduces the Doppler factor to $13.5$. 

The projected length of the \extendedjet{} in X-rays is about $10\,$kpc, while the widths of the jet is about $1\,$kpc (see Fig. 2 in \cite{kwt13}). Since the radio and X-ray emission rapidly fade beyond the second bend at $\sim 12\,$arcsec of the jet, we do not take any emission from beyond that point into account. 

Owing to the sparse data set of the \extendedjet, the parameters are less well constrained than for the core component. The photon index of the jet X-ray spectrum, $\Gamma_X = 1.8\pm 0.1$, results in a spectral index of the SED of $\alpha = 0.2\pm 0.1$ (if defined as $\nu F_{\nu}\propto \nu^{\alpha}$). This implies an IC origin of the X-ray emission. On these scales the only plausible external soft photon field is the CMB, while the synchrotron radiation is too weak to induce a significant amount of SSC emission. In the Thomson limit the spectral index of IC/CMB radiation follows the same dependence on the electron spectral index as the synchrotron spectral index ($s = 3-2\alpha$), and so the electron spectral index is given by $s = 2.6 \pm 0.2$. 

The minimum electron Lorentz factor is restricted to values below $\sim 40$, or else the jet X-ray spectrum would not be fitted. There is no observational constraint on the maximum electron Lorentz factor. Klein-Nishina effects are not important for the IC/CMB process as long as $\gamma_{max}\lesssim 10^8$. 

At $\gamma$-rays the poor spatial resolution means that it is not possible to discriminate between core and extended emission. Since the core parameters fail to reproduce the VHE measurement, we invoke the \extendedjet. Using the IC/CMB model we can infer constraints on more parameters. Since the IC/CMB spectrum is determined by the electron distribution, we can use the observed VHE spectrum to derive the maximum electron Lorentz factor. In the Thomson limit the observed scattered photon energy depends on the characteristic CMB photon energy following $\nu_{ic,max} = 4\delta_b\Gamma_b\nu_{cmb}\gamma_{max}^2$. With $\nu_{max} \sim 10^{27}\,$Hz we find $\gamma_{max}\sim 5\times 10^6$.

Extrapolating the jet X-ray spectrum to higher energies implies an observed break frequency of $\sim 10^{25}\,$Hz. This gives $\gamma_{br} \sim 5\times 10^5$, which is related to the escape time scale parameter $\eta_{esc}$ by Eq. (\ref{eq:gammab}). Assuming that the cooling is dominated by IC/CMB cooling, we find $\eta_{esc} \sim 15$. This implies that electrons reside in any given subunit of the jet for only a small amount of time.

The assumption of dominating IC/CMB cooling is well justified given that the Compton dominance $CD$ is much larger than unity. Taking the ratio of the measured X-ray and radio fluxes of the \extendedjet, the $CD$ is at least $30$. The $CD$ can be estimated by the ratio of the IC/CMB and synchrotron energy densities, giving $CD = 1.62\times 10^{-11} (\Gamma_b/B_0)^2$. Hence, the magnetic field is smaller than $\sim 7.3\,\mu$G. However, the X-ray flux is produced by electrons at the low-energy end of the electron distribution, while the radio flux is from electrons with higher energies. Thus, the real $CD$ should be taken at different frequencies, which are related to the same electron Lorentz factor. Owing to the lack of other measurements of the \extendedjet{} we cannot give a tighter constraint, but the $CD$ could be about a factor of $10$ larger than the ratio of the X-ray and radio fluxes. Estimating the $CD$ as the ratio of the respective energy densities is only a rough approximation, since it does not take into account the different beaming properties of synchrotron radiation and IC radiation of external fields (\cite{d95}).

The remaining parameters have been chosen to allow the modeling of both the resolved spectra and the VHE measurement, and are given in the third column of Table \ref{tab:inputcom}. The resulting model is presented in Fig. \ref{fig:extmodel} as the solid lines, giving an excellent fit to the observed fluxes of the jet and the VHE measurement. The model will be discussed extensively in section \ref{sec:dis}.

The cooling time of the highest energetic electrons in the extended jet is on the order of $850\,$years. This is a factor of $300$ below the light travel time along the \extendedjet, and demonstrates the need for constant re-acceleration of the electrons on scales smaller than the widths of the jet. 

Since the medium in the extended components is much thinner than in the \blob, synchrotron self-absorption can be neglected entirely in the large-scale jet following Eq. (\ref{eq:nussa}).

The combined fit of the \blob{} and the entire jet gives a satisfactory representation of the data of AP Librae with the major implication that the \extendedjet{} is responsible for the VHE regime. 

\subsubsection{Comparison to multizone models} \label{sec:comphbs}

Recently, \cite{hbs15} presented their model to explain the spectrum of AP Librae. They developed a multizone model, which is similar to our modeling in some ways. 

However, their explanation of the VHE emission differs substantially from our model. Their electron distribution function has not been derived from first principles, which sets the break $\Delta s$ of the slope of the electron distribution function (see Eqs. (\ref{eq:fastcooling}) and (\ref{eq:slowcooling})), but was assumed ad hoc to take up arbitrary values. This allowed the electron distribution to be extended to $\gamma_{max} = 4\times 10^{6}$ using a steep spectrum after the break electron Lorentz factor. Thus, they were able to keep the blob IC emission below the Fermi level and the high-energy end of the synchrotron component below the X-ray measurement, while explaining the VHE flux with emission from the one-zone blob. Owing to the self-consistent approach, we have no freedom for $\Delta s$, and cannot raise $\gamma_{max}$ beyond $\sim 10^4$ in the blob. 

So far, the data does not lead to a firm conclusion whether to decide if the core or the large-scale jet is responsible for the VHE emission. In section \ref{sec:ext} we present several ways to test our scenario, and compare our interpretation to other BL Lac objects with extended jet emission in section \ref{sec:morph}. These arguments support the view that the VHE emission originates in the extended jet.

%
%
\section{Discussion} \label{sec:dis}
\begin{table}
\caption{Resulting parameters for the individual components using the input parameters of Table \ref{tab:inputcom}.}
\begin{tabular}{lccccc}
				  & Blob 		& pc-jet 		& Extended jet \\
	\hline
	$\delta_b$ 	 	  & $17.8$		& $17.8$	 	& $13.5$ \\
	$n_e$ [cm$^{-3}$] 	  & $2.9\times 10^{2}$	& $4.5\times 10^{-7}$ 	& $1.6\times 10^{-9}$ \\
	$\gamma_{min}$ 		  & $1.6\times 10^2$ 	& $2.0\times 10^2$ 	& $4.0\times 10^1$ \\
	$\gamma_{br}$ 		  & $9.1\times 10^3$ 	& $7.7\times 10^3$ 	& $6.2\times 10^5$ \\
	$\gamma_{max}$ 		  & $1.0\times 10^4$ 	& $1.0\times 10^4$ 	& $5.0\times 10^6$ \\
	$s$ 			  & $2.0$ 		& $2.0$ 		& $2.6$ \\
	cooling regime		  & slow		& slow			& slow \\
	main cooling 		  & SSC			& SYN			& IC/CMB \\
	$u_{e}$ [erg/cm$^3$]	  & $1.6\times 10^{-1}$	& $3.0\times 10^{-10}$	& $1.4\times 10^{-13}$ \\
	$u_{p}$ [erg/cm$^3$]	  & $4.3\times 10^{-1}$	& $6.8\times 10^{-10}$	& $2.5\times 10^{-12}$ \\
	$u_{B}$ [erg/cm$^3$]	  & $6.4\times 10^{-5}$	& $9.9\times 10^{-7}$ 	& $2.1\times 10^{-13}$ \\
	$u_{rad}$ [erg/cm$^3$]	  & $3.1\times 10^{-3}$	& $6.5\times 10^{-12}$	& $9.9\times 10^{-17}$ \\
	$u_{DT}$ [erg/cm$^3$] 	  & $1.1\times 10^{-7}$ & -		 	& - \\
	$P_{j}$ [erg/s]		  & $4.5\times 10^{44}$	& $1.8\times 10^{46}$	& $1.3\times 10^{46}$ \\
	$\epsilon_{eq,e}$	  & $3.8\times 10^{-4}$	& $3.3\times 10^{3}$ 	& $1.5$ \\
	$l_{jet}$ [pc]		  & - 			& $286.5$		& $1.4\times 10^{5}$ \\
	$N_{jet}$ 		  & $1$			& $214$			& $215$ \\
	\hline
\end{tabular}
\newline
\footnotesize{$\delta_b$: Doppler factor; $n_e$: electron density; $\gamma_{min}$, $\gamma_{br}$, and $\gamma_{max}$: minimum, break, and maximum electron Lorentz factor, respectively; $s$ is the electron spectral index below the break; $u_{e}$: energy density in relativistic electrons; $u_{p}$: energy density in (cold) protons; $u_{B}$: magnetic energy density; $u_{rad}$: total radiative energy density; $u_{iso}$: energy density of the isotropic field in the galactic frame; $P_j$: total jet power in the observer's frame; $\epsilon_{eq,e}$: equipartition parameter between magnetic field and electrons; $l_{jet}$: de-projected length of the jets; $N_{jet}$: number of subunits. } 
\label{tab:resultcom}
\end{table}
%
%
\subsection{Particle population and power requirements} \label{sec:ppe}
The parameters of the electron distribution derived by the code are listed in Table \ref{tab:resultcom}. In all cases the cooling operates in the slow mode, which gives a break in the electron distribution of unity. In combination with the canonical value of $s = 2.0$, this restricts the \blob{} maximum Lorentz factors to $10^{4}$. 

The electron densities decrease outward, as is expected for an expanding jet. In order to reach quasi-neutrality of the plasma, we assume one proton per radiating electron. As is common in leptonic models, these protons are assumed to be cold and do not contribute to any emission process. As can be seen in Table \ref{tab:resultcom}, the protons dominate the particle energy density in all cases.

The photon energy density $u_{rad}$ emerging from the emission site is calculated from the model SEDs as
\begin{eqnarray}
 u_{rad} = \frac{4d_L^2}{cR_b^2\delta_b^4 N_{jet}} \int \nu F_{\nu} \td{\ln{(\nu)}} \label{eq:urad} .
\end{eqnarray}
The resulting values are also listed in Table \ref{tab:resultcom}. In each component the radiation energy density is several orders of magnitude below the particle energy density, and only in the blob does it exceed the magnetic energy density. Hence, AP Librae's jet is radiatively inefficient.

The accretion disk luminosity is on the order of a few times $10^{-3}$ times the Eddington luminosity, which is $L_{edd} = 3.75\times 10^{46}\,$erg/s in the case of AP Librae. The total jet power in each case is calculated by
\begin{eqnarray}
 P_{j} = \pi R_b^2 \Gamma^2 c N_{jet} \sum_i u_i \label{eq:jetpower},
\end{eqnarray}
where the sum includes the energy densities in electrons, protons, magnetic field, and radiation. The entire jet obeys the Eddington limit. This is noteworthy for the \extendedjet, as we discuss in section \ref{sec:ext}.

It seems counterintuitive that the electrons in the jet should exhibit higher energies than in the \blob{} close to the initial acceleration region. However, our jet model requires continuous re-acceleration of the particles throughout the jet. Since the maximum Lorentz factor is restricted by the equality of the acceleration and the cooling rate, and the cooling rate drops farther out in the jet due to the weaker magnetic field and less intense soft radiation fields, the maximum Lorentz factor would increase with increasing distance from the core, if the acceleration rate does not drop as strongly as the cooling rate. This is reasonable, since the cooling and acceleration rates have different dependencies on the underlying parameters, such as the magnetic field (e.g.,\cite{ba12}). We will elaborate more on the high electron Lorentz factors in section \ref{sec:morph}.

In many blazar modeling attempts, equipartition between the magnetic field and the radiating particles is assumed. This reduces the free parameters, and results in the lowest energy budget required to fit the data. We have not taken this into account, and basically all components are far away from equipartition. The equipartition parameters is defined as $\epsilon_{eq,e} = u_B/u_e$. The \blob{} is particle dominated, while the \pcjet{} and the \extendedjet{} are magnetic field dominated. We do not include the cold proton population in this value, because they do not contribute to the emission processes. If we included the protons in the equipartition parameter, the ratios would decrease, and the \extendedjet{} would be considered particle dominated. Interestingly, the \extendedjet{} is very close to equipartition, and this condition can be realized without a drastic change in the parameters. The equipartition magnetic field would result in $2.2\,\mu$G, while the other parameters would remain unchanged. 

%
\subsection{Extended jet} \label{sec:ext}
The IC/CMB modeling of the \extendedjet{} reproduces the observations. The IC/CMB model has been used in many sources with extended jets, but has several potentially problematic implications (\cite{hk06}).  

First of all, it requires a decent amount of beaming, which implies a relativistic flow on kpc-scales coupled with a small viewing angle. Unfortunately, dynamic range limits preclude any detection of movement in the large-scale jets inhibiting the determination of the bulk Lorentz factors. The small viewing angle results in very large physical sizes of the jets on the order of Mpc scales. Even though some radio galaxies with such large jets have been observed, a larger number of misaligned sources should have been observed for any more closely aligned large-scale jet. In our modeling we only consider the part of the jet that is upstream of the bend, and has an angular size of roughly $10\,$arcsec. This corresponds to about $10\,$kpc at the distance of AP Librae. Owing to the small viewing angle of only $4^{\circ}$ the de-projected length of the jet becomes $\sim 140\,$kpc. Considering that the jet continues for about $10\,$arcsec beyond the bend, AP Librae's jet appears to extend up to $300\,$kpc, rivaling the largest radio galaxies. 

However, there are several arguments in favor of strong beaming in the case of AP Librae. The particularly sharp bend in the radio and X-ray maps at $\sim 10\,$arcsec could be due to a low angle between the jet and the line of sight. At small viewing angles a small intrinsic bend of the jet is significantly enhanced in the observer's frame and in the projection on the sky. Since the X-ray flux strongly decreases beyond the bend (cf. Fig. 2 in \cite{kwt13}), we interpret this as a reduction in the observed IC flux due to reduced beaming. The synchrotron emission is not as influenced by this beaming effect as the IC emission (\cite{d95}). More sensitive mapping at radio and X-ray frequencies could test this prediction.

Furthermore, the counter-jet has not been detected in radio maps on arcsec and milliarcsec scales or in the X-ray map. Since jets should emerge on both sides of the accretion disk, the counter jet must be much weaker than the observed jet. Under the assumption that both jets are intrinsically identical, the counter-jet must be strongly de-beamed. Since far away from the black hole there is still no sign of the counter-jet, we can further conclude that even at great distances the jet must exhibit relativistic speeds without significant deceleration. Otherwise, there should be signs of the counter-jet at great distances from the black hole.

Since the electrons emitting in the X-ray band are less energetic than the radio emitting electrons, the power in relativistic particles is high and in some cases exceeds the Eddington luminosity even without accounting for cold protons. Our modeling suggests that in AP Librae the Eddingtion luminosity constraint holds for the large-scale jet even with the dominating contribution of cold protons.

Additionally, in the few sources where the synchrotron component of the extended jet is well constrained by detections up to the optical and/or UV band, the IC/CMB model usually predicts a measurable and steady flux in the Fermi band. Using this prediction, \cite{mg14} and \cite{mea15} ruled out the IC/CMB model for the quasars 3C 273 and PKS 0637-752. An inversion of this test could be useful for AP Librae. As implied by our modeling, the Fermi and H.E.S.S. data can be satisfactorily explained by the IC/CMB model. Owing to the required highly relativistic electrons, the jet is expected to emit synchrotron emission up to the UV band. Since the contribution of the galaxy in the UV band is much smaller than in the optical and IR bands, the jet should be detectable despite its proximity to the bright core. If the jet is not detected in the UV with an upper limit below the predicted flux of our model curve, the IC/CMB is ruled out as the origin of the VHE emission because the required electron energies cannot be matched. 

The HE and VHE domain could be used to implicitly test our model. In the model presented here, the VHE emission is produced in the extended jet. Owing to the large spatial extent, the light crossing time is large, and the emitted flux is not expected to change on time scales shorter than several thousand years. The observed flux in the energy range dominated by jet emission is therefore expected not to drop below the level reproduced by the model. A reassuring argument is the relatively stable lightcurve recorded with the Fermi satellite. Apart from the flare in 2013, which lasted for several months, no significant variability has been recorded and the lightcurve is well fit by a constant (\cite{hea15}). Unfortunately, the low flux of the flare of AP Librae in the high-energy domain does not allow detailed spectroscopic studies. The spectrum integrated over the entire flare extends only up to 10 GeV. This is consistent with the model described in this paper. Below $\sim 30\,$GeV the emission is dominated by IC scattering originating in the compact \blob, and is thus subject to variations on short time scales. A temporal increase in the electron density in the blob can easily account for the variation. Unfortunately, the sparse multiwavelength data set inhibits the construction of a meaningful model. At energies higher than $\sim 30\,$GeV the emission is dominated by the extended jet and no changes in the total flux are expected.

An increase in the flux at VHE energies during a flare might still originate from an additional flaring component and would thus not rule out our model. However, this component should also reveal itself at lower energies by changing the spectrum in the UV and potentially in the X-ray band because of the higher $\gamma_{max}$ required for such a flare.

On the other hand, the flux above $\sim 30\,$GeV should not drop below the observed level, because this could not be explained by emission processes in the \extendedjet. More observations on this source by the H.E.S.S. experiment and the future Cherenkov Telescope Array are strongly encouraged.

%
\subsection{Scenario for the jet structure in AP Librae} \label{sec:mod}
The preceding results and discussions are combined in a scenario that reproduces the observations.

This scenario includes a continuous jet being launched by the central engine. The densities in these extended zones are very low, and the electrons mostly produce synchrotron radiation through interactions with the ambient magnetic field. Only the large size of the \extendedjet{} of more than $140\,$kpc (de-projected) allows for significant IC emission produced by interactions with the CMB. The electrons in the jet are continuously reaccelerated throughout the jet. This may be achieved in turbulence producing numerous acceleration regions on scales of a few percent of the jet widths, which are too small to be resolved by radio or X-ray observations conserving the observed homogeneity. The acceleration process itself could be a combination of shocks, reconnection or shear layer acceleration, depending on the microscopic properties of the turbulent regions within the large-scale structure. For simplicity, we assume that the initial electron energy distribution is preserved as in the model of \cite{bk79}. 

Occasionally, an overdensity like the \blob{} considered in section \ref{sec:fit} is launched, which may be more magnetized than the continuous jet medium. The blob moves outward, expands, and dissolves into the surrounding jet medium. Owing to the initial higher densities, the blob emits X-rays and HE $\gamma$-rays through IC interactions with the synchrotron, accretion disk, and isotropic radiation fields. The last two become less important as the blob moves outward. 

This scenario explains the observational features. Most importantly, the MOJAVE observations show that blobs emerge from the central engine, but in most cases cannot be followed over more than a few months. Only three components were identified in more than one observation, which allowed the derivation of the apparent speed of the jet (\cite{lea13}). If this is not related to dynamic range limits, this could be interpreted as blobs dissolving into the jet medium. 

%
\subsection{Comparison to BL Lac objects with extended X-ray jets} \label{sec:morph}
\begin{table*}
\caption{Properties of BL Lac objects with extended X-ray jets.}
\begin{tabular}{lccccccl}
	Source		& $z$	  & $\Gamma_X$ 	& $B$ [$\mu$G] 	& $\gamma_{min}$	& $\gamma_{max}$	& $s$	& Ref. \\
	\hline
	PKS 0521-365 	& $0.055$ & $2.4$	& $160$		& - 			& - 			& - 	& [1] \\
	OJ 287 		& $0.306$ & $1.7$	& $4.9$		& $40$			& - 			& $2.4$ & [2] \\
	AP Librae	& $0.049$ & $1.8$	& $2.3$		& $40$ 			& $5.0\times 10^6$	& $2.6$	& [3], this work \\
	3C 371		& $0.051$ & $\sim 2$	& $73$		& $1$ 			& $1.0\times 10^8$ 	& $2.35$& [4] \\
	S5 2007+777	& $0.342$ & $\sim 1$	& $2.5$		& $70$			& $3.0\times 10^5$ 	& $2.5$	& [5] \\
	PKS 2201+044	& $0.027$ & $\sim 2$	& $<100$	& -			& -			& - 	& [4] \\
	\hline
\end{tabular}
\newline
$z$ is the redshift; $\Gamma_X$ is the X-ray photon index; $B$ the magnetic field; $\gamma_{min}$ and $\gamma_{max}$ the minimum and maximum electron Lorentz factor, respectively; $s$ the electron spectral index; the references are [1] \cite{bea02}, [2]  \cite{mj11}, [3] \cite{kwt13}, [4] \cite{sea07}, and [5] \cite{sea08}.
\label{tab:xjets}
\end{table*}
Currently, six BL Lac objects are listed in the X-JET database. We present some of their properties in Table \ref{tab:xjets}. 

The sources do not exhibit a clear trend concerning the X-ray emission process, because for only three sources (OJ 287, AP Librae, S5 2007+77) the photon indices clearly imply an IC origin. The other sources have been modeled with synchrotron emission. The radiation process is independent of redshift. There might be a correlation to the magnetic field strength in the extended jets, since the IC/CMB dominated jets require a smaller magnetic field ($<10\,\mu$G) than the synchrotron dominated ones ($>50\,\mu$G).

Comparing our extended jet model, constrained by the VHE $\gamma$-ray measurement, to the parameters used for OJ 287 (\cite{mj11}) and S5 2007+777 (\cite{sea08}), similar values with respect to magnetic field, electron distribution, and the respective powers are found. The magnetic fields are on the order of a few $\mu$G, the minimum electron Lorentz factors on the order of $\sim 50$, and the electron spectral indices are $\sim 2.5$ (as listed in Table \ref{tab:xjets}). The powers in magnetic field and relativistic particles are on the order of a few $10^{45}$erg/s. Optical and/or infrared observations on OJ 287 and S5 2007+777 have not revealed the jets. With the resulting upper limits, the maximum electron Lorentz factor can be constrained to a few times $10^5$, and thus VHE emission is not expected from the extended jets of these sources. 

As mentioned in section \ref{sec:ppe}, we require a maximum electron Lorentz factor of $5\times 10^6$, much higher than the maximum Lorentz factor of the core ($\sim 10^4$). However, considering the sources listed in Table \ref{tab:xjets}, such high Lorentz factors in extended jets seem to be common. Especially in the sources where the X-ray emission is considered to be of synchrotron origin, a maximum Lorentz factor on the order of $\sim 10^8$ is required (\cite{sea07}). Hence, extended jets must be efficient particle accelerators. 

%
%
\section{Conclusions} \label{sec:con}
We have modeled the broad-band spectrum of the LBL source AP Librae, which is unusual both for its broad-band $\gamma$-ray emission and for its extended X-ray jet. Owing to the spectral features in the X-ray and $\gamma$-ray emission the electron distribution function is strongly constrained. This rules out an explanation of the SED by a single-zone model. The failure of the one-zone model is confirmed by quantitative modeling, as has also been noted by other authors (\cite{tea10,hbs15}). Since the maximum of the IC component is located in the energy range between $100\,$keV and $100\,$MeV (the Swift-BAT and the Fermi-LAT measurements), the maximum electron Lorentz factor is restricted to a low value ($\sim 10^4$), which is inadequate to explain the VHE emission. Even the addition of observationally unconstrained external fields, such as the accretion disk or the dusty torus, cannot account for the VHE flux. For standard values of the electron spectral index of $s\sim 2$, the hard spectrum in the X-rays further constrains the minimum electron Lorentz factor to greater than $10^2$, which is inadequate to model parts of the radio and infrared spectrum. Instead, it is demonstrated that emission from the extended jet may be responsible for many parts of the SED. The synchrotron component is due to a superposition of the one-zone blob, and the jet on pc-scales. This model is supported by VLBI observations of the MOJAVE program, which resolves the jet on these scales.

For the VHE emission we have invoked the \extendedjet{} on arcsec scales. Considering IC scattering of the CMB in a uniform jet resulted in a successful fit of the data. The parameters are reasonable with minimum electron Lorentz factor $\gamma_{min} = 40$, electron spectral index $s=2.6$, and magnetic field $B_0 = 2.3\,\mu$G. The success of our model depends critically on the value of the maximum electron Lorentz factor. If electrons are accelerated to $\gamma_{max} = 5\times 10^{6}$ throughout the jet, the IC/CMB emission contributes strongly in the VHE regime.

The parameters of the \extendedjet{} model match those of modeling attempts of the sources OJ 287 (\cite{mj11}) and S5 2007+777 (\cite{sea08}). These sources are also BL Lac objects with an extended, IC dominated X-ray jet. The inferred parameters, especially the values of the magnetic field on the order of a few $\mu$G and the parameters of the electron distribution ($\gamma_{min}\sim 50$, $s\sim 2.5$), are similar to our parameters. Thus, extended jets must be efficient accelerators on all scales. since the entire volume of the jet is required to produce the emission (\cite{bdf08}). 

We have also presented explicit tests to confirm the extended jet origin of the VHE emission in AP Librae. This includes a direct detection of the jet in the UV band, and an indirect confirmation if the flux at VHE $\gamma$-rays does not drop below the average level. This confirmation is important, because the emission of VHE radiation by an extended blazar jet would be an extraordinary result, which might even have consequences for the acceleration of ultra high-energy cosmic rays. Further observations of AP Librae in all energy bands is strongly encouraged. 

If other jets exhibit extended VHE emission, these sources could be resolved by the future Cherenkov Telescope Array (CTA) owing to its superior angular resolution compared to the current generation of Cherenkov experiments.

%
%
\begin{acknowledgement}

The authors wish to thank Markus B\"ottcher for the numerical code and fruitful discussions. We also thank G. Bicknell and S. Kaufmann for discussions. We thank the anonymous referee for a constructive report, which helped to clarify the presentation.
This work was supported by the German Ministry for Education and Research (BMBF) through Verbundforschung Astroteilchenphysik grant 05A11VH2. 
This research has made use of data from the MOJAVE database that is maintained by the MOJAVE team (\cite{lea09}). 
This paper is based on observations obtained with Planck (http://www.esa.int/Planck), an ESA science mission with instruments and contributions directly funded by ESA Member States, NASA, and Canada.

\end{acknowledgement}

\section*{Note added in proof}
While this paper was undergoing the referee process, we became aware of a publication by Sanchez, D.A., et al. (2015, MNRAS 454, 3229), which also interprets the TeV emission as originating in the large-scale jet. These authors use a one-zone model for the jet emission, which gives different results and parameters than our extended jet model.
%
%

%

%
%
\begin{appendix}
\section{Numerical code} \label{app:code}
Models presented in this study have been obtained by an extension of the leptonic, self-consistent, steady-state, one-zone code of \cite{bea13}.
%
\subsection{Blazar code}
The code evaluates the spectrum of a homogeneous, spherical emission region. From the input parameters, the electron distribution function is derived self-consistently as a broken power law. The break energy is calculated from the competition between electron cooling and electron escape, giving
\begin{eqnarray}
 \gamma_{br} = \frac{1}{t_{esc}\kappa_{cool}} = \frac{c}{\eta_{esc}R \kappa_{cool}} \label{eq:gammab}.
\end{eqnarray}
Here, $c$ is the speed of light, $R$ the source radius, and $\eta_{esc}$ a free parameter to scale the escape time in multiples of the light crossing time $R/c$. The electron cooling term $\dot{\gamma} = \kappa_{cool}\gamma^2$ (where $\gamma$ is the electron Lorentz factor) takes into account both synchrotron and inverse Compton cooling. The latter depends on the strengths of the different seed photon fields. The seed fields considered are synchrotron photons (with the density depending on the strength of the magnetic field $B_0$), photons of the cosmic microwave background (CMB), photons from the accretion disk, and components that reflect the accretion disk radiation into the jet such as the dusty torus. These depend on the specific conditions and constraints derived for the source. 

Starting from an electron distribution function following a simple power law of the form
\begin{eqnarray}
 n(\gamma) = n_0 \gamma^{-s}, \quad \gamma_{min}<\gamma<\gamma_{max} \label{eq:ngammainput}
\end{eqnarray}
the code evaluates the break energy $\gamma_{br}$ and constructs the broken power law, taking into account two forms of the electron distribution depending on the fast and slow cooling regime, respectively:
\begin{eqnarray}
 n^{fast}(\gamma_{br}<\gamma<\gamma_{min}) &=& n_0  \gamma^{-2} \nonumber \\
 n^{fast}(\gamma_{min}<\gamma<\gamma_{max}) &=& n_0  \gamma^{-(s+1)} \label{eq:fastcooling} \\
 n^{slow}(\gamma_{min}<\gamma<\gamma_{br}) &=& n_0 \gamma^{-s} \nonumber \\
 n^{slow}(\gamma_{br}<\gamma<\gamma_{max}) &=& n_0 \gamma^{-(s+1)} \label{eq:slowcooling} .
\end{eqnarray}
The normalization $n_0$ is derived from the equation
\begin{eqnarray}
 L_{inj} = \pi R_b^2 \Gamma_b^2 c m_ec^2 \int\td{\gamma} \gamma n(\gamma) \label{eq:electronnormalization},
\end{eqnarray}
with the electron injection luminosity $L_{inj}$, the bulk Lorentz factor of the emission blob $\Gamma_b$, and the electron rest mass energy $m_ec^2$. These parameters are either input parameters or constants.

The spectrum of the accretion disk is calculated from the accretion disk luminosity $L_{disk}$ and the mass of the black hole following the standard Shakura-Sunyaev type (\cite{ss73}). Apart from the direct thermal disk emission the IC scattered emission of the photon field emitted by the disk is computed. This depends sensitively on the injection height of the emission blob $z_0$ above the black hole (\cite{ds94}).

The thermal accretion disk radiation is likely to give rise to further thermal emission fields such as radiation from a BLR or reprocessed emission from a thick dusty torus. While the direct emission of such hypothetical components are low (non-detections providing upper limits) IC scattered emission might be more significant owing to a favorable geometry and relativistic beaming. The reprocessed component is defined by a temperature $T_{DT}$ and an energy density $u_{DT}$. If this component scatters a fraction $\tau_{DT}$ of the accretion disk radiation, and is bounded by an outer radius $r_{DT}$, the energy density can be estimated as
\begin{eqnarray}
 u_{DT} = \frac{L_D \tau_{DT}}{4\pi r_{DT}^2 c} \label{eq:ublr} .
\end{eqnarray}

The spectra are calculated in the blob frame, and afterwards transformed to the observer's frame by applying the cosmological distance correction and the Lorentz boost with the Doppler factor $\delta_b = [\Gamma_b(1-\beta \cos{\vartheta_{obs}})]^{-1}$, where $\beta = v_b/c = (1-\Gamma_b^{-2})^{1/2}$ is the normalized speed of the blob, and $\vartheta_{obs}$ is the angle between the jet direction and the line of sight.

Further details of the code are given in \cite{bea13,bdf08,bc02,bb00,bms97}.

\subsection{Modeling extended components}
Since in AP Librae the jet is resolved in both radio and X-ray observations, the parameters of the extended jet can be derived. We modified the code, which allows us to model the extended jet as the sum of many identical subunits. While it is certainly a simplification to assume that the extended jet will be homogeneous throughout its volume, using the same average parameters for the entire jet reflects limited spatial resolution.

The number of subunits in the extended jet is calculated from the ratio of the jet volume versus the volume of one subunit:
\begin{eqnarray}
 N_{jet} \sim \left[ \frac{\pi R_{jet}^2 l_{jet}}{\frac{4}{3}\pi R_{jet}^3} \right] \label{eq:ejnumber}.
\end{eqnarray}
Here, we set the radius of a subunit equal to the observable radius of the jet $R_{jet}$. The de-projected length of the extended, cylindrical jet is $l_{jet} = l_{jet}^{\prime}/\sin{\vartheta_{obs}}$. Correspondingly, $l_{jet}^{\prime}$ is the projected length of the jet, which is the observed quantity. 

The code then evaluates the spectrum of one component and by multiplying this result with $N_{jet}$, obtains the spectrum of the entire jet. This implicitly assumes that the blobs are non-interacting. Since the radiation densities in each blob are extremely low (only the combination of all blobs gives a measurable emission), this assumption is well justified.

%
%
\section{Constraining the one-zone model} 
\subsection{Parameters of the high $\gamma_{min}$ model} \label{app:ssc}
The high minimum Lorentz factor for the one-zone model implies a relatively low magnetic field in order to match the flux of component 0. Here, we demonstrate that the high minimum Lorentz factor is well constrained by the X-ray spectrum, if the electron spectral index follows the canonical value of $s\sim 2$.

The SSC emission by the blob component of the $\nu F_{\nu}$ spectrum (in units of erg/cm$^2$/s) is given by
\begin{eqnarray}
 \nu_s F_{\nu_s} &=& \, \frac{R_b^3\sigma_T m_e c^3}{4 d_L^2} \delta_b^2 (1+z) \epsilon_s^2 \nonumber \\
 &\times& \intl_{\gamma_{min}}^{\infty} \td{\gamma} \frac{n(\gamma)}{\gamma^2} \intl_0^{\infty} \td{\epsilon} n_{syn}(\epsilon) G(q,\Gamma_k) \label{eq:nuFnussc}
\end{eqnarray}
with the Thomson cross section $\sigma_T = 6.65\times 10^{-25}\,\mbox{cm}^2$, the electron rest mass $m_e = 9.1\times 10^{-28}\,\mbox{g}$, the normalized scattered photon energy $\epsilon_s = h\nu_s/m_ec^2$, the normalized synchrotron photon energy $\epsilon = h\nu/m_ec^2$, the synchrotron photon density $n_{syn}(\epsilon)$, and the Klein-Nishina cross section kernel
\begin{eqnarray}
 G(q,\Gamma_k) &=& 2q\ln{q} + (1+2q)(1-q) + \frac{\Gamma_k^2q^2(1-q)}{2(1+\Gamma_kq)} \\
 q &=& \frac{\epsilon_s}{\Gamma_k(\gamma-\epsilon_s)} \\
 \Gamma_k &=& 4\epsilon\gamma \label{eq:KNparam} .
\end{eqnarray}
The inverse Compton kinematics require $\gamma>\gamma_{min}$ with
\begin{eqnarray}
 \gamma_{min} = \frac{\epsilon_s}{2} \left[ 1+\sqrt{1+\frac{1}{\epsilon\epsilon_s}} \right] .
\end{eqnarray}

The above equations hold for isotropic electrons and synchrotron photons, and include the full Klein-Nishina cross section. Eq. (\ref{eq:nuFnussc}) folds the electron distribution with the synchrotron photon distribution weighted by the cross section. Thus, the resulting spectrum will be curved because it results from multiple broken power laws. 

Even though a single electron with Lorentz factor $\gamma_e$ emits a broad spectrum of synchrotron photon energies, most photons exhibit the characteristic frequency (in the observer's frame)
\begin{eqnarray}
 \nu = 4.18\times 10^6 \delta_b b \gamma_e^2 \, \mbox{Hz} \label{eq:synfreq}.
\end{eqnarray}
Similarly, the characteristic SSC frequency emitted by an electron with Lorentz factor $\gamma_s$ can be estimated by
\begin{eqnarray}
 \nu_{s} = 4\nu \gamma_s^2 \, \mbox{Hz} \label{eq:sscfreq}.
\end{eqnarray}
One should be aware that in the SSC process any electron can scatter any synchrotron photon. Hence, $\gamma_s$ in Eq. (\ref{eq:sscfreq}) is in most cases not equal to $\gamma_e$ used in Eq. (\ref{eq:synfreq}).

Analyzing the synchrotron frequencies for the parameters given in the first column of Table \ref{tab:inputcom} and \ref{tab:resultcom}, one finds 
\begin{eqnarray}
 \nu_{min} &=& 7.6\times 10^{10}\, \mbox{Hz} \\
 \nu_{br} &=& 2.5\times 10^{14}\, \mbox{Hz} \\
 \nu_{max} &=& 3.0\times 10^{14}\, \mbox{Hz},
\end{eqnarray}
where each frequency corresponds to $\gamma_e = \{ \gamma_{min}, \gamma_{br}, \gamma_{max} \}$, respectively. We use the parameters $b=0.04$ and $\delta_b = 17.8$. In principle, $\nu_{br}$ corresponds to the maximum in the synchrotron component, while the synchrotron flux cuts off beyond $\nu_{max}$. However, the close proximity of these two frequencies does not allow us to discriminate between them in Fig. \ref{fig:ozssc}. 

\begin{figure}
\centering
\includegraphics[width=0.48\textwidth]{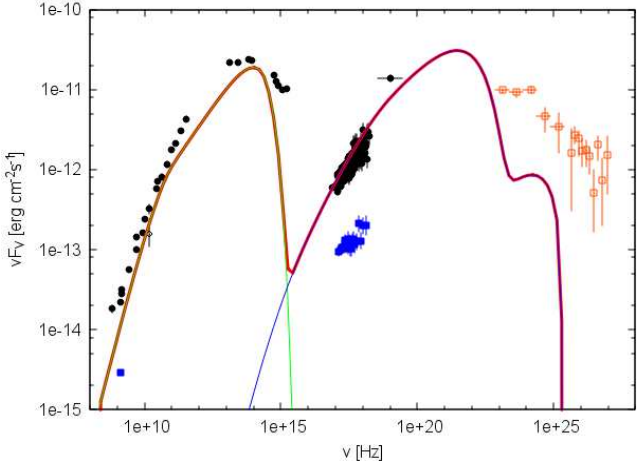}
\caption{Same as Fig. \ref{fig:ozssc}, but with $\gamma_{min}=100$.}
\label{fig:oztest}
\end{figure}

Scattering by electrons with Lorentz factors close to $\gamma_{min}$ will have a significant effect on the SSC spectrum, since most electrons exhibit these energies. Hence,
\begin{eqnarray*}
 4\nu_{min}\gamma_{min}^2 &=& 8.0\times 10^{15}\, \mbox{Hz} \\
 4\nu_{br}\gamma_{min}^2 &=& 2.6\times 10^{19}\, \mbox{Hz} \\
 4\nu_{max}\gamma_{min}^2 &=& 3.1 \times 10^{19}\, \mbox{Hz}.
\end{eqnarray*}
In fact, around each of these frequencies the SSC spectrum becomes softer. For a smaller value of $\gamma_{min}$ the softening sets in at a lower frequency. Fig. \ref{fig:oztest} illustrates this case, where the minimum Lorentz factor is reduced to $\gamma_{min}=100$ while keeping all other parameters as in Table \ref{tab:inputcom}. In this case $4\nu_{max}\gamma_{min}^2 = 1.2\times 10^{19}\,$Hz, and the model curve does not fit the Swift-BAT point. 

The cut-off SSC frequency becomes
\begin{eqnarray*}
 4\nu_{max}\gamma_{max}^2 &= 1.2\times 10^{23}\, \mbox{Hz}.
\end{eqnarray*}
Since $\nu_{max}$ is beyond the maximum of the synchrotron component, the cut-off SSC frequency is also beyond the maximum of the SSC component. It is used to constrain $\gamma_{max}$, since the SSC component should not exceed the Fermi measurement.

These considerations apply only to IC scattering in the Thomson limit. This can be checked using Eq. (\ref{eq:KNparam}). One obtains for the maximum Lorentz factor and normalized synchrotron energy
\begin{eqnarray*}
 \Gamma_k &=& 4\left( \frac{h\nu_{max}}{\delta_b m_ec^2} \right) \gamma_{max} \\
 &=& 0.005.
\end{eqnarray*}
Since the Klein-Nishina regime requires $\Gamma_k>1$, the first-order SSC component is in the Thomson regime. 

On the other hand, the second-order SSC component is in the Klein-Nishina regime, which can be deduced from the hard, non-exponential cut-off. The inverse Compton kinematics in the Klein-Nishina domain require a cut-off at normalized energy $\epsilon_{ssc,max} = \gamma_{max}$. For $\gamma_{max}=1.0\times 10^4$ , this corresponds to a maximum observed frequency of $\nu_{ssc,max}= 2.2\times 10^{25}\,$Hz. Indeed, the second-order SSC component cuts off at this frequency.

\subsection{Parameters of the hard electron distribution model} \label{app:ssch}
\begin{figure}
\centering
{\includegraphics[width=0.48\textwidth]{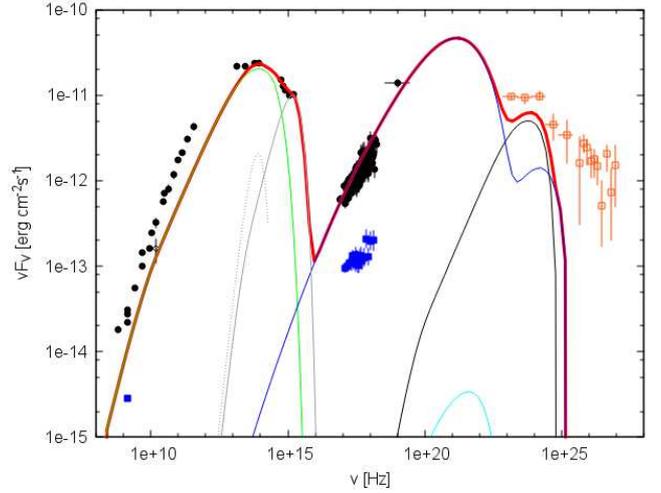}}
\caption{One-zone model with hard electron distribution. Data points are as in Fig. \ref{fig:spec}. Line codes are as in Fig. \ref{fig:ozssc} and \ref{fig:ozmodel}.}
\label{fig:ozhard}
\end{figure}

Here we demonstrate that a similar fit to that in Fig. \ref{fig:ozmodel} can be achieved for a lower $\gamma_{min}$, but with a much harder electron distribution. The parameters are listed in Table \ref{tab:inputhard}, and the resulting fit is shown in Fig. \ref{fig:ozhard}.
\begin{table}
\caption{Input parameters for the hard electron distribution model.}
\begin{tabular}{lc}
	$L_{inj}$ [erg/s] 	  & $2.0\times 10^{44}$ \\
	$\gamma_{min}$ 		  & $30$ 		\\
	$\gamma_{max}$ 		  & $8.0\times 10^3$ 	\\
	$s$ 			  & $1.5$ 		\\
	$\eta_{esc}$ 		  & $100$		\\
	$B_{0}$ [G] 		  & $0.08$ 		\\
	$R$ [cm] 		  & $4.0\times 10^{15}$ \\
	$z_{0}$ [pc] 		  & $0.3$ 		\\
	$\Gamma_{b}$ 		  & $10$ 		\\
	$\vartheta_{obs}$ [deg]	  & $2.0$ 		\\
	$L_d$ [erg/s] 		  & $1.3\times 10^{44}$	\\
	$\tau_{DT}$ 		  & $0.01$		\\
	$r_{DT}$ [pc] 		  & $0.95$		\\
	$T_{DT}$ [K] 		  & $1.0\times 10^{3}$	\\
\end{tabular}
\label{tab:inputhard}
\end{table}

The hard electron distribution model faces the same problem as the high $\gamma_{min}$ model. Both are difficult to realize by current acceleration theories and cannot explain the VHE emission, because of the low value of $\gamma_{max}$.

Consequently, the choice of specific sets of parameters for the one-zone model does not affect the difficulty of explaining emission in the TeV range. This paper focuses on the interpretation that the \extendedjet{} emits the VHE radiation.

\end{appendix}
\end{document}